\documentclass{ws-procs9x6}
\begin{document}
\title{Polarized sea measurements at JPARC}
\author{M.~Contalbrigo, A.~Drago and P.~Lenisa}
\address{Universit\`a di Ferrara and INFN, \\
44100 Ferrara, Italy}

\maketitle

\abstracts{
Large double spin-asymmetries can be foreseen for Drell-Yan 
production in $p p$ scattering at JPARC energies. The sign of the asymmetries
can be used to discriminate between different model calculations of sea quark
distributions.}

\section{Introduction}

Helicity distributions of valence quarks have been investigated in many 
experiments and are now well known. On the other hand, the distributions
of the polarized sea are poorly known. In particular, concerning 
$\Delta \bar u(x)$, fits assuming a totally flavour symmetric sea suggest a
sign opposite respect to fits based on a flavour broken sea \cite{Gluck}.

The situation is even more confused concerning the transversity polarized
sea. While the estimate of the numerical value of $\delta u(x)$ is
more or less similar in all model calculations (the discrepancy is a factor
2 roughly), the uncertainties in the theoretical predictions for
$\delta \bar u(x)$ are much more significant. Actually, there are essentially
only two model calculations, one based on the Chiral Color-Dielectric Model
(CCDM) \cite{Barone} and one on the so-called 
Chiral Quark-Soliton Model (CQSM)
\cite{Wakamatsu}. The two predictions differ not only on the size of the
polarized sea, but also on the sign. While an experiment based on 
Drell-Yan production in $p p$ scattering could only extract with large 
error bars the transversity polarized quark distributions, the sign of the
asymmetry could be determined without any ambiguity, as long as the 
valence region is tested ($x_B>0.1$). An experiment at $\sqrt s \sim $ 10 GeV
as foreseen at JPARC can explore this region.

\section{Double spin asymmetries in Drell-Yan production at JPARC}

\paragraph{Longitudinal case}

The helicity distributions have been estimated using
the two different scenarios discussed in Ref.~[1]:

\begin{itemize}
\item{valence scenario:\\
\begin{equation}
\frac{\Delta \bar d(x,\mu)}{\Delta \bar u(x,\mu)}=
\frac{\Delta u(x,\mu)}{\Delta d(x,\mu)}; 
\Delta s(x,\mu)= \Delta \bar s(x,\mu) = 0.
\end{equation}}
\item{standard scenario:\\
\begin{equation}
\Delta \bar u(x,\mu)=\Delta u_{sea}(x,\mu)= 
\Delta \bar d(x,\mu)=\Delta d_{sea}(x,\mu)=
\Delta \bar s(x,\mu)=\Delta s(x,\mu)  
\end{equation}}
\end{itemize}

The longitudinal double spin asymmetry ($A_{LL}$) in Drell-Yan production 
has then been evaluated for $\sqrt s \sim 10$ GeV 
corresponding to the c.m. energy 
of the fixed target experiments foreseen at JPARC:

\begin{equation}
A_{LL} =
\hat{a}_{LL}\frac{\sum_q e_q^2
  [\Delta q(x_1) \Delta \bar q(x_2) + \Delta \bar q(x_1) \Delta q(x_2)]} 
{\sum_q e_q^2
  [q(x_1)\bar q(x_2) + \bar q(x_1) q(x_2)]}
\end{equation}

\begin{figure}[htb]

\centerline{\epsfxsize=4.1in\epsfbox{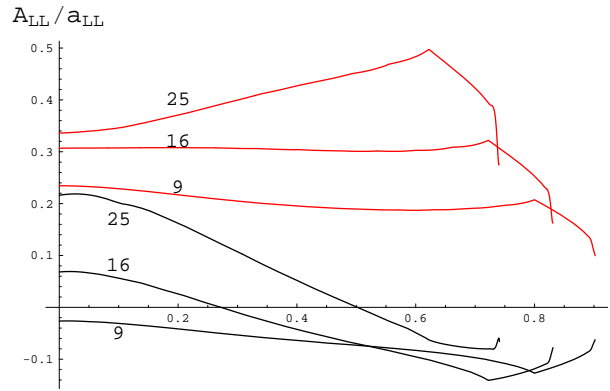}}   
\vspace{-7cm}
\caption{Longitudinal double spin asymmetries for Drell-Yan production
at JPARC energies for different values of $Q^2$ as a function of $x_F$. 
The upper curves are related to the evolution of the so called 
valence scenario of Ref.~[1], 
while the lower ones to the standard scenario.\label{figure1}}
\end{figure}

The results are reported in Fig.~\ref{figure1}. The asymmetries foreseen at
JPARC energy are much larger than the ones predicted for 
RHIC with $\sqrt s \sim 200 GeV$ 
where a polarized Drell-Yan experiment has also been proposed~\cite{Barone4}.
It is worthwhile to remark
the different sign of the asymmetry 
in the two scenarios at 
high values of $x_F$. It is also interesting to remark that also 
the CCDM and the CQSM both predict a positive sign.

\paragraph{Transverse case}

The evolution of the transversity distributions has been evaluated starting 
from two different assumptions:

\begin{equation}
\delta \bar u(x,\mu) = \Delta \bar u(x,\mu) 
\end{equation}
consistent with the CCDM in Ref.~[2], or
\begin{equation}
\delta \bar u(x,\mu) = - \Delta \bar u(x,\mu) 
\end{equation}
consistent with the CQSM in Ref.~[3]. 

The asymmetry in Drell-Yan production for the case of double transverse
polarization ($A_{TT}$) has then been estimated: \footnote{Note that the 
asymmetry is dominated by the $\delta \bar u$ distribution.}

\begin{equation}
A_{TT} =
 \hat{a}_{TT}\frac{\sum_q e_q^2
  [\delta q(x_1) \delta \bar q(x_2) + \delta \bar q(x_1) \delta q(x_2)]} 
{\sum_q e_q^2
  [q(x_1)\bar q(x_2) + \bar q(x_1) q(x_2)]}
\end{equation}

\begin{figure}[htb]

\centerline{\epsfxsize=4.1in\epsfbox{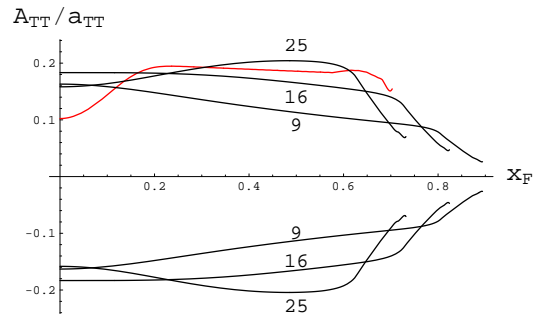}}   
\vspace{-7cm}
\caption{Transverse double spin asymmetries for Drell-Yan production at
JPARC energies for different values of $Q^2$ as a function of $x_F$. 
The upper curves correspond
to the evolution starting from the assumption consistent with the CCDM, 
while the lower ones with the one consistent with the CQSM. 
The thin line in the upper part is a prevision of
the CCDM. \label{figure2}}
\end{figure}

The asymmetry is reported in Fig.~\ref{figure2}. 
Also in this case the asymmetries are predicted to be considerably larger than
those foreseen at RHIC.
It is remarkable to notice
how the different assumptions at the model scale, reflect themselves in the
sign of the asymmetry giving a chance to directly discriminate between
the models.

\section{Discussion and Conclusions}

It must be remarked that the differences between the previsions of the two 
models for the transversity distribution should 
actually be attributed not only to the differences between the two
models, but also to the different technique adopted in the evaluation of the
antiquark distributions. In Ref.~[2] the matrix element defining
the antiquark distribution has been evaluated by an explicit insertion
of 4q intermediate states. In Ref.~[3], use has been made of an
analytic continuation to negative values of $x$, a procedure that is 
probably unsafe \cite{Jaffe,Barone2}. \footnote{Notice that a discrepancy
between the previsions of the two calculations shows up also for the 
longitudinaly polarized sea, as discussed in Ref.~[7].}

A polarized Drell-Yan experiment testing both $A_{LL}$ and $A_{TT}$ could provide
a clear answer to the problem of computing antiquark distributions in
quark models, a question which is of paramount importance in the theoretical
calculations.

\end{document}